# A Framework for Digital Asset Risks with Insurance Applications

JUNE | 2024

# A Framework for Digital Asset Risks with Insurance Applications

**AUTHORS** Zhengming Li, PhD Candidate

Jianxi Su, PhD, FSA

Maochao Xu, PhD

Jimmy Yuen, FSA, FCIA

**SPONSOR** General Insurance Research Committee

Casualty Actuarial Society

Actuarial Innovation and Technology Strategic Research Program Steering Committee

SOA Research Institute Emerging Issues Pool

Caveat and Disclaimer

The opinions expressed and conclusions reached by the authors are their own and do not represent any official position or opinion of the Society of Actuaries Research Institute, the Society of Actuaries or its members. The Society of Actuaries Research Institute makes no representation or warranty to the accuracy of the information.

# CONTENTS

# A Framework for Digital Asset Risks with Insurance Applications

## Executive Summary

The remarkable growth of digital assets, starting from the inception of Bitcoin in 2009 into a $1 trillion market in 2024, underscores the momentum behind disruptive technologies and the global appetite for digital assets. 2024 is a particularly pivotal year for the ecosystem as it marks the adoption by broader financial markets via the Bitcoin ETF, BlackRock's deployment of the BUIDL fund on the Ethereum blockchain and the anticipation of further institutional follow-ons.

Until now, institutional participants have mainly focused on either adopting blockchain technology into their own processes via private permissioned blockchains or outright investment into major digital assets like Bitcoin. While this exposes organizations to the technology, it does not allow them to capitalize or prepare for the bulk of major advancements. This report offers an alternative path for participation that more appropriately aligns corporate interests with grassroots developments: methodical underwriting of digital asset risk on public blockchains.

What distinguishes digital assets from traditional financial instruments is its governance by smart contracts encoded in blockchain protocols. These instruments promise increased efficiency, transparency, and financial inclusion but also attract a spectrum of threats, including cybercriminals aiming to exploit vulnerabilities for financial gain. The often-decentralized structure of protocols and absence of central authority render digital assets vulnerable to various forms of exploitation, including hacking, manipulation, and disruptions, as evidenced by numerous past incidents. Recognizing the critical need to bolster cyber-resilience, digital asset insurance has emerged as a pivotal risk management tool. While demand for digital asset insurance is rising, the market remains nascent, characterized by limited coverage and understanding of systemic cyber risks. Peer-to-peer insurance platforms and innovative business models are filling the void left by traditional insurers, but challenges persist in assessing and pricing cyber risks.

This project develops a framework to enhance actuaries' understanding of the cyber risks associated with the developing digital asset ecosystem, as well as their measurement methods in the context of digital asset insurance. By integrating actuarial perspectives, we aim to enhance understanding and modeling of cyber risks at both the micro and systemic levels. The qualitative examination sheds light on blockchain technology and its associated risks, while our quantitative framework offers a rigorous approach to modeling cyber risks in digital asset insurance portfolios. This multifaceted approach serves three primary objectives: i) offer a clear and accessible education on the evolving digital asset ecosystem and the diverse spectrum of cyber risks it entails; ii) develop a scientifically rigorous framework for quantifying cyber risks in the digital asset ecosystem; iii) provide practical applications, including pricing strategies and tail risk management. Particularly, we develop frequency-severity models based on real loss data for pricing cyber risks in digit assets and utilize Monte Carlo simulation to estimate the tail risks, offering practical insights for risk management strategies. As digital assets continue to reshape finance, our work serves as a foundational step towards safeguarding the integrity and stability of this rapidly evolving landscape.

## Section 1: Introduction

Since 2009, digital asset ecosystems[1] have undergone explosive growth, radically changing our ability and mediums of transacting. The digital asset market cap alone has grown from $237B in 2020 to $1.18T. Digital assets like Bitcoin exist on a publicly viewable and immutable ledger, accessible to anyone, anywhere in the world with an internet connection. This is driven by smart contracts, or protocols embedded directly in the blockchain, which automate and secure a variety of financial services such as payments, lending, borrowing, and investing. As these digital protocols replicate the familiar structures of financial services, they offer a glimpse into a future where finance is more open, accessible, and connected than ever before.

The latest growth cycle in this decade has been catalyzed by primitives in decentralized finance ("DeFi"), digital collectibles and gaming finding product market fit. These coinciding successes have led to continued iterations and a rapid influx of global talent.

However, the ecosystem is increasingly attractive to not only enterprising builders, but also a broad spectrum of threats from cyber criminals, who aim to steal funds, manipulate prices, or disrupt the normal operations of digital asset platforms (Caporale et al., 2021). The decentralized nature and lack of central authority of blockchain technology compounds the growth of cyber risks. Malicious actions can pose severe financial consequences for individuals and businesses holding digital assets. For example, in 2022, an attack on the Axie Infinity Ronin Bridge, a platform which enabled users to deposit and retrieve assets in the eponymous game, resulted in the theft of nearly $600 million worth of digital assets. While some of the funds lost due to malicious activities may eventually be recovered and returned to the owners, they are the minority. These incidents serve as a reminder that cyber risks are omnipresent in the digital asset space and need to be consistently managed.

As the digital asset ecosystem continues to grow and evolve, maintaining its cyber-resilience[2] has become a top priority. To this end, insurance has naturally emerged as a tool of interest for protecting digital asset users from potential losses due to malicious activities. Demand for digital asset insurance[3] has steadily increased and whilst traditional insurers have been absent in developing their own solutions, new players and business models have flourished to administer coverage needs. Most notably, there has been development of peer-to-peer (a.k.a., P2P) insurance platforms or "DeFi insurers" to protect user digital assets. However, the digital asset insurance market is still early, and a lack of sophisticated understanding of the cyber risks involved presents significant obstacles to its maturation (Kwock et al., 2022). In addition, the digital asset ecosystem is characterized by its code-sharing practices, which often involve developers using and building upon each other's code repositories (Chen et al., 2021). While the practice can promote innovation and the advancement of new technologies, it also gives rise to a highly interconnected ecosystem where cyber risks may become highly correlated among various protocols (Zhou et al., 2022). These interconnections inevitably lead to the accumulation of systemic risks throughout the insurance portfolios of digital asset insurers, which can place a significant strain on their solvency (Kwock et al., 2022). Consequently, current digital asset insurance products are limited in their coverage, with significant risks often excluded and low coverage limits. To advance the digital asset insurance market, a more sophisticated understanding of the cyber risks in the digital asset ecosystem is urgently called upon.

[1] Digital asset refers to a type of asset that uses advanced cryptography to secure transactions and control the creation of new units on a distributed ledger. Bitcoin and Ethereum are among the most well-known digital assets.

[2] In the context of this paper, cyber-resilience refers to a digital asset service provider's or user's ability to withstand and/or recover from cyber exploits.

[3] In the context of this paper, digital asset insurance refers to a type of insurance that is designed to protect users accessing protocols on public blockchains from losses of digital assets due to malicious exploits.

In this report, we marry academic rigor and practicing actuaries' working experience to fill the knowledge gap. There are some relevant studies of cyber risks to protocols in the literature, but mainly from a security perspective in the domain of computer science. For instance, Danielius et al. (2020) analyzed some of the specific risks of smart contracts in the Ethereum blockchain electricity market. Wan et al. (2021) proposed a mixture of qualitative and quantitative methods to comprehend practitioners' perceptions and practices of smart contract security. Lee et al. (2023) examined various attacks against bridges in the blockchain ecosystem and suggested several mitigation strategies to address most of these attacks. Zhou et al. (2022) comprehensively evaluated DeFi attacks and incidents by introducing a shared reference frame and exploring potential defensive mechanisms. Despite the research surrounding smart contract security, there is a noticeable absence of studies dedicated to the actuarial modeling of smart contract security risk. It is worth mentioning that there is literature loosely related to our project, which focuses on modeling cyber risks from an abstract network perspective. For example, Sun et al. (2021) developed a frequency-severity actuarial model for aggregated enterprise-level breach data to inform ratemaking and underwriting in insurance. Similarly, Sun et al. (2023) discussed a multivariate frequency-severity framework for healthcare data breaches, utilizing a vine copula approach to model dependence among the number of affected individuals at the state level. For a recent review of cyber risk modeling in insurance and actuarial science, please refer to He et al. (2024).

This report contains a comprehensive qualitative and quantitative analysis of the cyber risks in the digital asset ecosystem at the micro level, incorporating insurance perspectives. Our analysis will not only consider the cyber risks associated with individual protocols, but also examine the systemic risks present in theoretical digital asset insurance carrier portfolios. The qualitative part of our paper will provide the actuarial community with a clear and concise education about the blockchain technology underlying the digital asset ecosystem and its associated limitations and risks. Meanwhile, the quantitative part will be the first significant effort from the actuarial literature to develop a scientifically rigorous yet practically relevant framework for modeling different types of cyber risks inherent in a complex digital asset insurance portfolio. The quantitative framework we establish will have immediate applications in academia and practice, including but not limited to pricing, risk capital analysis, and understanding the economic role of insurance in maintaining the cyber-resilience of the digital asset ecosystem.

The rest of the report is organized as follows. Section 2 provides a qualitative understanding of the cyber risks involved in the context of the digital asset insurance business. Section 3 introduces the real loss data of digital assets and performs exploratory data analysis. In Section 4, we detail the development of a frequency-severity approach to model the real loss data of digital assets. Section 5 illustrates the application of our proposed model in insurance pricing and tail risk estimation. In Section 6, we summarize our findings and offer insights for further discussion.

## Section 2: The Risk Framework

This section aims to develop the readers' qualitative understanding of the cyber risks involved in the context of the digital asset insurance business. It is noteworthy that it is not our intention – nor possible – to cover all the risks associated with digital asset services. Instead, our focus is an introduction to risks that actuaries working with digital asset-related insurance products would typically prioritize for monitoring and analysis.

Through a careful analysis of past incidents and attack vectors, we make recommendations on viable covered perils to advance the industry.

## 2.1 GLOSSARY OF TERMS

Let us begin with a glossary to address key terms used in the report.

**Audit** – Code analysis performed by a third party to identify potentially fatal flaws and logic.

**Blockchain** – Reference to the distributed ledger that cryptographically links blocks of transactions together.

**DeFi ("Decentralized Finance") –** Broadly refers to financial applications on public blockchains.

**Digital Assets** – Often used interchangeably with crypto or tokens in reference to the digital representation of an asset (existing solely on the blockchain or otherwise).

**Flash Loan** – A unique, uncollateralized loan that allows users to borrow without collateral and pay back loans within the same blockchain transaction.

**Off-chain** – Reference to activity that occurs outside of the blockchain.

**On-chain** – Reference to activity occurring on the blockchain.

**Oracle –** A source of truth and information on the blockchain.

**Private Key** – A cryptographically generated key used to decrypt messages and is connected to a Public Key, which is viewable. Private keys are typically heavily guarded.

**Protocol –** A group of smart contracts that provides the functionality of a codified company. It is sometimes referred to as a blockchain application.

**Re-entrancy** – A common vulnerability where a function can be interrupted and called repetitively before execution.

**Rugpull** – Refers to the action of a project team overtly stealing invested funds.

**Smart Contracts** – Programs that are published to and executed on the blockchain.

**TVL ("Total Value Locked")** – The total amount of assets stored inside the protocol and/or ecosystem.

**Wallet** – A tool that allows blockchain users to custody their own assets and gain greater access to a greater array of blockchain applications.

**Web2** – The current state of the internet as we know it.

**Web3** – The future state of the internet, which will be connected by blockchain ecosystems and its applications, granting real ownership for end users.

## 2.2 RISK CLASSIFICATION

Throughout this article, we treat blockchain protocols as entities akin to traditional companies, drawing parallels to deepen our understanding of their commercial and technological dimensions. Protocols are not just a technological construct, but commercial enterprises with vast economic incentives for founders and users.

By developing an awareness of where the potential pitfalls reside, we can methodically discern the frameworks best tailored to address them. Furthermore, this knowledge becomes instrumental in determining the most fitting risk transfer mechanisms to counter these threats, paving the way for a future where the Web3 industry is both innovative and secure for the everyday user.

We leverage empirical loss on the blockchain, which includes, amongst others, hacks by external actors and outright fraudulent behaviour by internal actors. As with all developing fields, there are insurable risks and uninsurable risks. This paper seeks to disaggregate the relevant risks.

To systematically assess their risks, we categorize them into four main areas: Business, Operational, Team and Technical factors.

Table 1

CATEGORIZATION OF RISKS IN THE DEFI ECOSYSTEM

| Category | Subcategory | Description |
|---|---|---|
| Business | Primitive | Borrowing and lending, exchanges, marketplaces, governance tools, etc. |
| | Regulatory | Abiding to compliance requirements. |
| | Blockchain | Risks of the chosen underlying blockchain(s) the protocol operates on. |
| | Novelty | The uniqueness and innovative aspects of the protocol that may not be battle-tested. |
| Operations | Response | Team's ability to respond to threats in a 24/7 operating environment. |
| | Market Volatility | The degree to which asset fluctuations that can trigger automated protocol responses. |
| | Asset | Types and quantities of digital assets are approved. |
| | User | Live interactions with the protocol are within the range of normalcy. |
| | Activation | Immediacy and process behind changes of significance to the protocol. |
| Team | Experience | The collective experience level of the team members. |
| | Anonymity | The degree to which team members' identities are disclosed. |
| | Centralization | Whether the team structure is centralized or decentralized and the change surface area of both groups. |
| Technical | Smart Contract | Vulnerabilities related to the code of smart contracts (re-entrancy, oracle manipulation, flash loans, etc.). |
| | Upgrade Authority | The ability of the team or external actors to make smart contract changes. |
| | Access Control | Measures in place to manage who can access certain information or funds to minimize single points of failure and human error. |
| | Signature | Use of single or multi-signature authorizations as appropriate. |
| | Integration | Inheritance of risks from smart contract or other service integrations. |
| | Phishing | Deceiving users or employees via compromised channels. |
| | Front-end | User interface vulnerabilities that could expose user data or funds. |

The adoption of blockchain technologies does not release teams from existing cyber best practices. Blockchain applications do not exist in isolation from existing technologies. Instead, they act as a complement to legacy technologies, merging them with the distinctive attributes of a decentralized public ledger. Understanding this mosaic is essential, as is the differentiation between genuine digital asset risks and legacy risks, including their corresponding attack vectors.

## 2.3 CYBER RISK STAKEHOLDERS

We start with an initial decomposition and definition of the risks faced by two distinct groups within the ecosystem:

1. Users – Individuals or groups of individuals actively interacting with the blockchain through purchasing, transferring, and investing in digital assets.
2. Protocol Operators – Teams comprised mainly of developers, community managers, product and business development professionals to drive the use of their application.

While other ecosystem stakeholders (e.g. investors) exist, we limit the scope of our study to these two groups to simplify explanations and recognize that parties may take on more expansive roles.

Based on historical losses, the risks faced by each group include:

- For Users
  - Phishing – Through targeted or broad phishing efforts, malicious actors attempt to trick users into revealing sensitive information. These efforts concentrate on luring users to click on links or download attachments to compromise their personal or financial information and gain access to their digital assets.
  - Front-end – Vulnerabilities associated with the user interface of an application, including links to insecure web pages.
  - Signing – Authorizing the use of user funds or permission to interact with an application leads to funds being drained from user accounts.
  - Smart Contract – Direct exploits on the smart contracts that have custody of user funds.
- For Protocol Developers
  - Back-end – Risks that revolve around server-side applications like database weakness or delayed communication between systems cause breaches.
  - Smart Contract – Vulnerabilities arising from exploitable code can lead to direct loss of funds through re-entrancy, oracle manipulation or other direct attacks – leading to team compensation or project abandonment.
  - Operational – Issues such as phishing on employees, rogue internal actors, system failure, human error, and poor key management may lead to exploitation.

While not exhaustive of all attack vectors, the issues highlighted above are representative of most exploits to date. Developers can create technically sound products, but it is necessary to also approach trust-building with end users as a non-technical endeavour – one that combines financial certainty and technological robustness. Users need to feel comfortable with the applications they are interacting with, and protocol teams need greater certainty that they are applying best-in-class risk management practices.

## 2.4 THE CURRENT DIGITAL ASSET INSURANCE MARKET

Digital asset insurance is in its infancy by many standards:

- Number of products that are available
- Scope of coverage
- Number of providers
- Coverage limits
- Value relative to other comparable decentralized finance services (e.g. borrowing)
- Number of dedicated professionals
- Projects funded to cover digital asset risk

Within the institutional space, demand for coverage outstrips the available capacity by a wide margin. The ratio of assets relative to insurance coverage is not only lower relative to a traditional custodian, but the scope of coverage is slimmer despite coverage being available as early as 2014. Recent failings of large, centralized exchanges further exacerbates this gap as opposed to drawing additional resources to support the industry. These are largely driven by the negative stigma associated with the industry – an important note as we delve into the insurable parts of the ecosystem.

Existing coverage offered to institutions resemble many existing products offered to “digital” organizations such as Cyber, D&O and E&O. Cold wallet (custody) and hot wallet crime coverage are also available, becoming increasingly understood by insurers and reinsurers.

Separately, as we look deeper into grassroots movements and organic development of risk transfer on the blockchain, lack of coverage is equally as pronounced. “DeFi Summer” and Covid-19 lockdowns were significant catalysts for the digital asset ecosystem, but users’ yield-seeking nature drove less attention and development in the critical financial security pillar. As of the writing of this article, the appetite for identifying the elements of insurability of blockchain-centric risks remain at an all-time low. Products that address grassroots users include smart contract, slashing and bridging; however, less than 1% of total assets deployed into the blockchain are covered. This is true no matter which public blockchain is being examined including Ethereum and upstarts such as Solana.

An interesting conundrum presents itself in this operating regime where *traditional organizations have high limits with low coverage while grassroots users are presented with low limits and wide coverage.*

This leads a self-insured model for most ecosystem participants with an illusion of protection. This is evident during loss events where there is limited recourse, and funds are typically irretrievable by participants.

## 2.5 LOSS LANDSCAPE

Looking into the history of loss events by entity type, three groups account for the majority of losses:

- Centralized exchanges – Licensed and regulated platforms that allow for the exchange of digital assets have lost $10B+ of user funds since 2022.
- Bridges – Infrastructure that allows for the transfer of digital assets between blockchains have cumulatively lost over $2B since 2022.

- Protocols – Financial applications built on top of blockchains and are effectual "companies" providing services to their users (including decentralized exchanges and lending protocols).

Within each of these entity groups, common loss causes include:

- Direct Exploit (e.g. re-entrancy, flash loan, etc.)
- Phishing (e.g. deceived employees or users)
- Internal malicious actors (e.g. crime)

Interestingly, many of the largest losses in the ecosystem's existence have been related to centralized exchange failure. That is, organizations that are involved with facilitating the trade of digital assets but regulated and operating with traditional processes and corporate structures. In fact, these organizations are able to misbehave because of the opacity of their processes and the incongruity between legacy technologies and the blockchain.

Contrast to centralized entities, decentralized counterparts have recorded comparatively lower losses. A vivid example is the operations of FTX. The company was found guilty of creating loans against customers and illiquid assets. It is more difficult to produce this type of falsified information on the blockchain, where transparency is a key feature.

Surprisingly, seemingly sophisticated firms, were the ones who caused the largest losses across the industry. Such issues, however, are not novel or unique to the blockchain realm. Traditional insurance companies have provisions and policies to cover such risks, whether it be through crime or D&O policies.

To determine what insurers may be interested in providing coverage for going forward, it's important to understand which events violate standard insurance principles and the ones that do not. Covering fraud and outright crime in which internal actors are acting in bad faith is clearly unacceptable. Similarly, phishing incidents, though sometimes covered under a traditional cyber policy, would be a difficult vector to defend against and fall under the company's own security awareness programs. Compounding this is the notorious difficulty in identifying the root cause of failure by examining the blockchain.

What separates direct exploits from other sources of loss is the novelty of this attack angle – namely being tied directly to blockchain smart contracts. On the blockchain, smart contracts hold assets, which are directly visible. It is thus possible to understand exactly what is being covered, identify where assets are being stored and trace the loss of funds and its interactions that led to loss. Unlike other vectors and traditional cyber policies where losses need to be quantified and adjusted, there is an explicit quantification that takes place in real-time. Thus, policies that focus on losses that are visible and traceable on the blockchain are directly leveraging one of the most powerful properties of this technology – a public ledger.

There is also an ecosystem of security professionals (code auditors, blockchain sleuths, bug bounty platforms) who can attest to the cause of an attack – often within minutes of its occurrence. By leveraging live updates from these services, insurers can gain immediate insight into the proximate cause of a loss and react accordingly. Thus, emergent risks tied to the blockchain itself are the risks insurers should focus on, as they are vital for the progression of the industry and a diversified source of exposure to traditional cyber risks.

## Section 3: The Exploit Database

The REKT Database[4] is one of the most comprehensive and public loss events databases available. Each data entry corresponds to a security event, documenting details of the occurrence date, the blockchain where the event occurred, the loss vector, and the amount of funds lost. Data considered in our report spans from January 2011 to December 2023, totaling 631 security incidents (filtered down from over 3,700 incidents). The incident subset represents events that have reasonable potential to be covered in an expanded cyber insurance policy (e.g., a direct exploit is covered, but not negligence or scams).

Readers should note that the smaller database relative to other lines of business is more than compensated for by its increased quality over traditional insurer loss datasets. The features of public blockchains mean loss data is complete, automatically captured, reported instantaneously and verifiable.

This section aims to provide readers with a comprehensive understanding of the evolving cybersecurity landscape by summarizing the frequency and severity of security incidents reported in the REKT Database.

### 3.1 SUMMARY OF INCIDENT FREQUENCY

Figure 1 displays the frequency of security incidents across various years. The data depicted highlights a noteworthy surge in reported loss events, notably peaking in 2022, a year when grassroots blockchain activity reached an all-time high. This escalation in security breaches underscores a growing concern for cybersecurity measures and emphasizes the pressing need for robust protective strategies.

Figure 1

THE NUMBER OF PUBLIC LOSS EVENTS IN THE REKT DATABASE

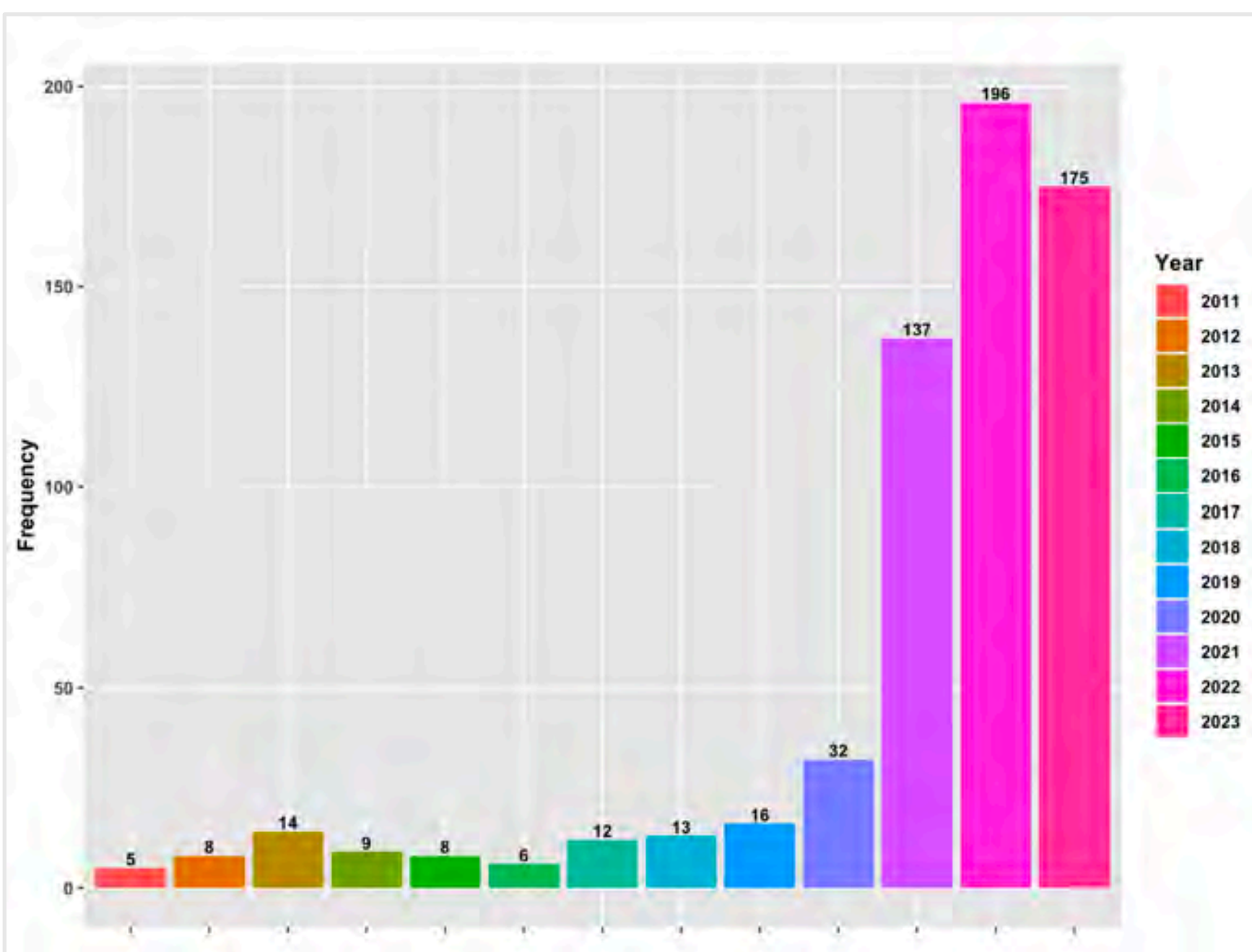

[4] Source: https://de.fi/rekt-database

Figure 2 displays the total assets of the DeFi ecosystem in USD from 2018 to 2023, with data sourced from DefiLlama[5]. The log transformation is applied to address the highly right-skewed nature of the data, which helps improve visualization. When this data is compared to the incident frequency shown in Figure 1, a clear positive relationship emerges between the growth of assets and the frequency of incidents. As the DeFi ecosystem grows, it also produces an unintended consequence: a larger target for attackers.

Figure 2

THE LOGARITHM OF TOTAL ASSETS IN THE DEFI ECOSYSTEM, MEASURED IN TERMS OF USD

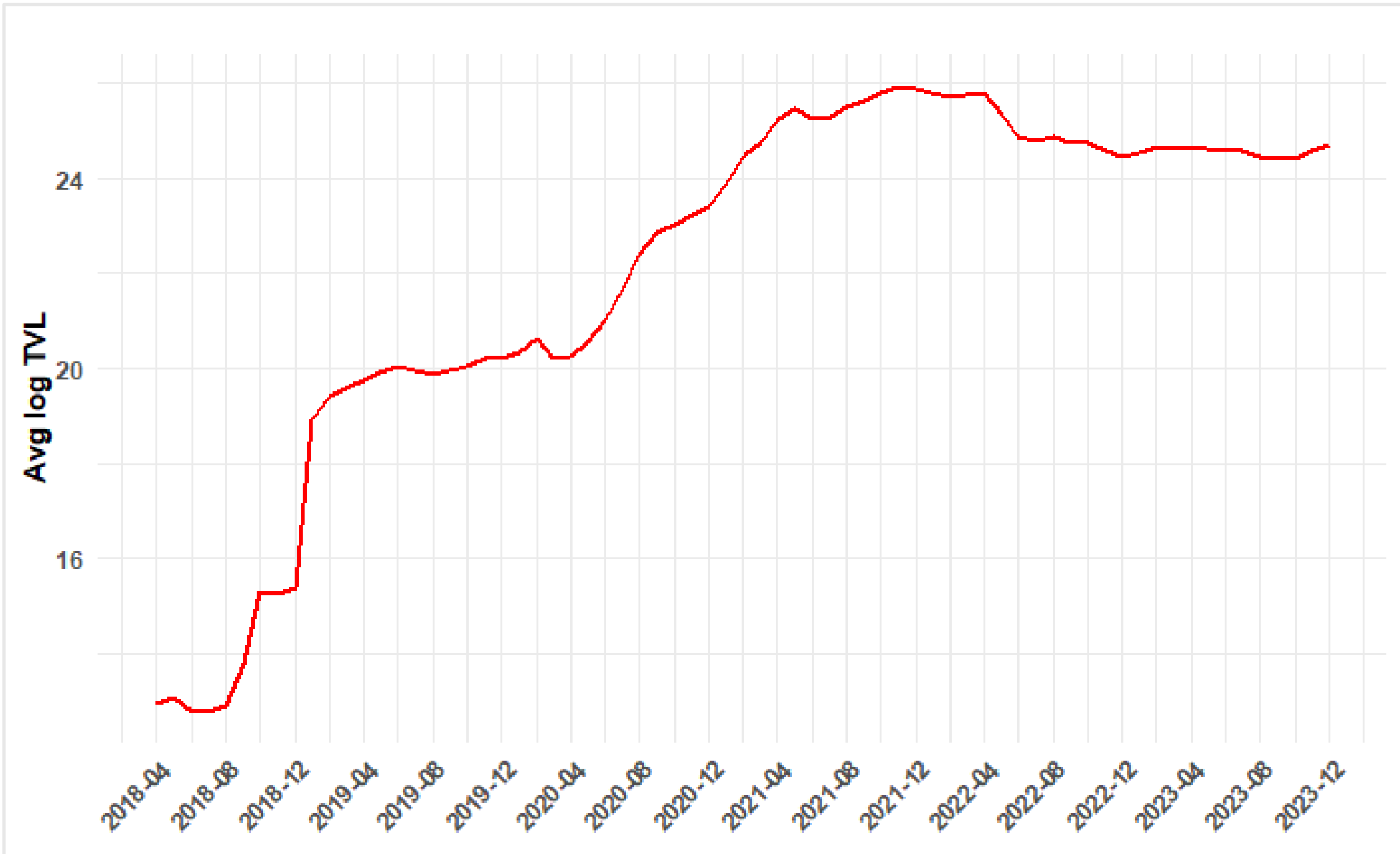

The loss dataset we used can be further decomposed into access control, flash loan attacks, Oracle issues, phishing, reentrancy, and direct subtypes. The distribution of the aforementioned subtypes is summarized in Figure 3. Notably, the "Other"[6] category records the highest frequency of incidents, followed by access control, whereas Oracle issues demonstrate the least occurrence. This breakdown sheds light on the prevalent types of security challenges insurers and stakeholders should consider when assessing risk factors and implementing preventive measures.

[5] Source: https://defillama.com/

[6] We use the “Other” category to be consistent with the REKT database, but the incidents in this category should be understood as “direct" exploits.

Figure 3

THE FREQUENCIES OF INCIDENTS RESULTING FROM VARIOUS TYPES OF ISSUES

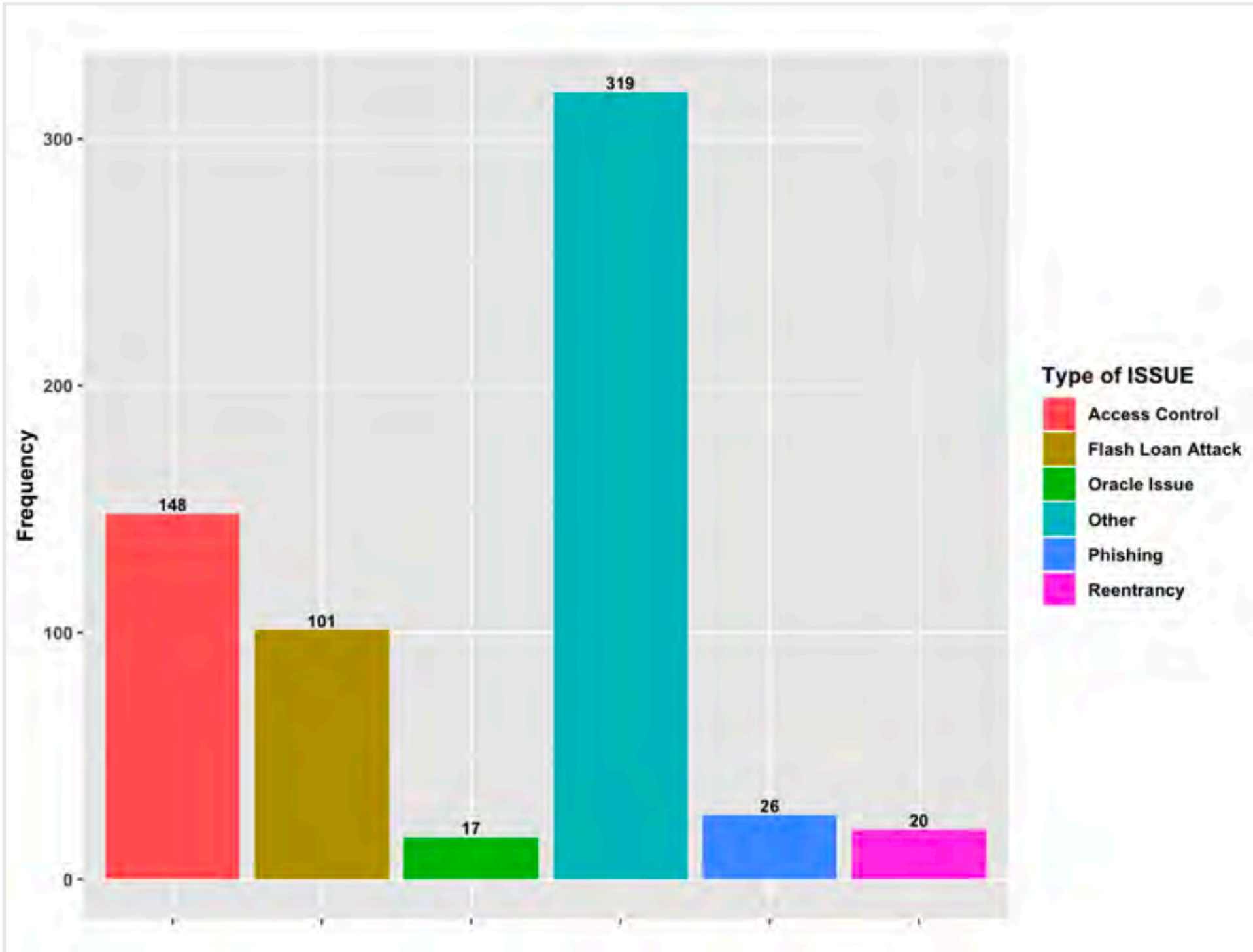

Figure 4 provides additional insight into the frequencies of issue types observed over the years. While overall exploit frequency has increased, the occurrences of Oracle manipulation, re-entrancy, and phishing have exhibited a stronger upward trajectory in recent years. Conversely, access control and other issue types declined slightly from 2022 to 2023.

Figure 4

THE FREQUENCIES OF INCIDENTS RESULTING FROM VARIOUS TYPES OF ISSUES OVER THE YEARS

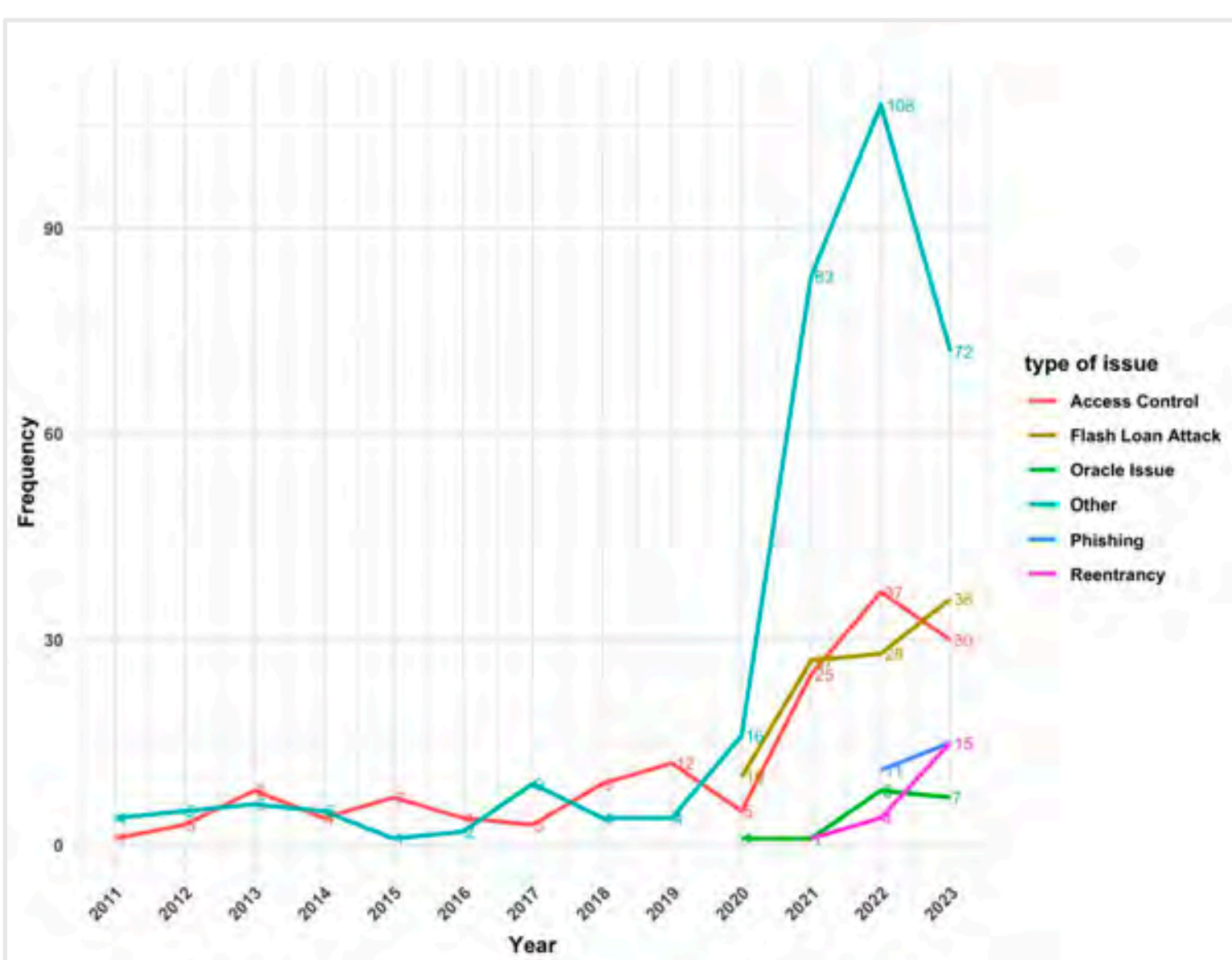

The left panel of Figure 5 displays the frequency of incidents for each blockchain over the years. Ethereum and Binance Smart Chain have shown a significant increase in incidents since 2018. In contrast, other blockchains like Solana have fewer recorded incidents, as summarized in the right panel, which illustrates trends across various chains since 2020. While some of these variations may stem from the number of protocols each blockchain supports, there is also some evidence that the inherent security of a blockchain's underlying technology and programming language plays a role. In addition, developers on newer blockchains are able to incorporate lessons from more established blockchains like Ethereum into their designs to mitigate common security risks.

However, it's important to remain objective. While new blockchains benefit from advancements in architecture that reduce certain risks, it's not without its challenges. Despite architectural improvements, new blockchains can face downtime and other new security risks that do not exist on battle-tested blockchains. This highlights an ever-evolving trade-off between innovative enhancements and network stability.

Figure 5

INCIDENT FREQUENCY FOR EACH CHAIN THROUGHOUT THE YEARS

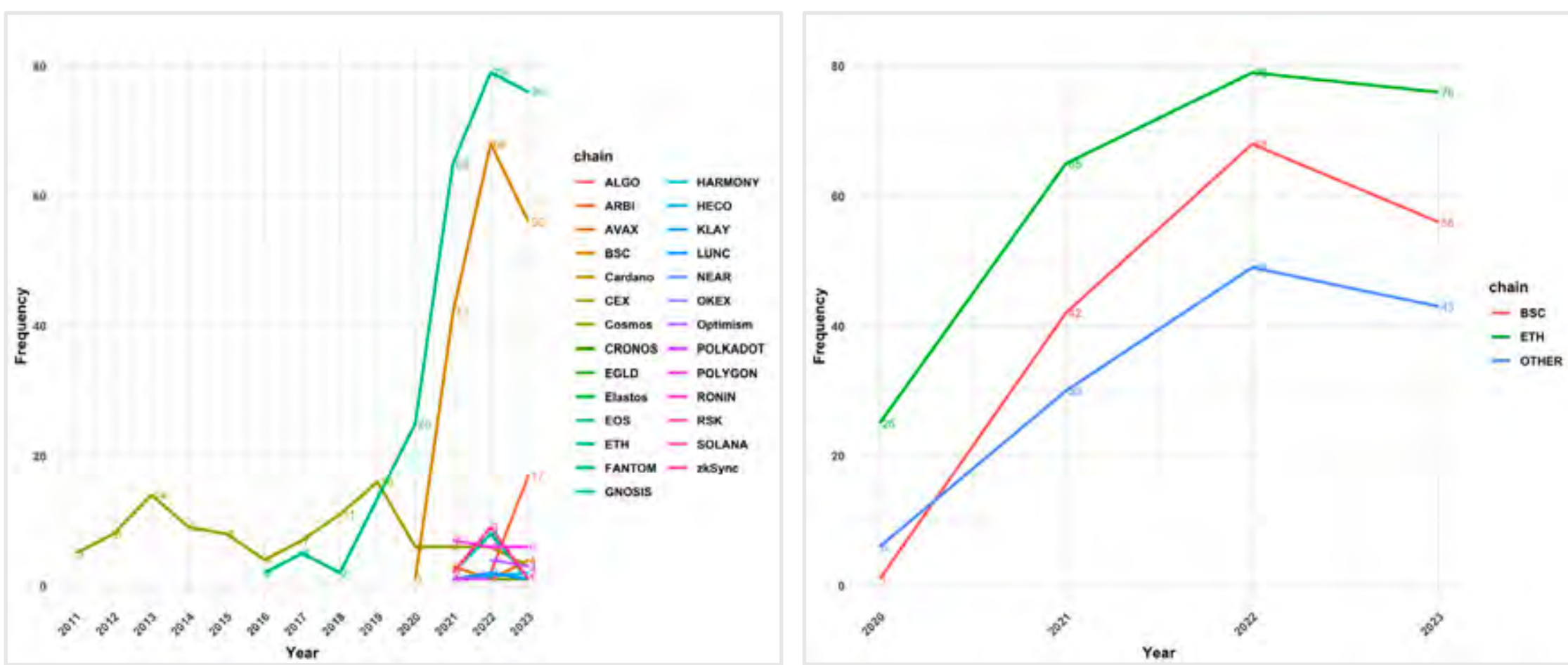

## 3.2 SUMMARY OF INCIDENT SEVERITY

Security incidents resulted in substantial financial losses for the broader digital asset ecosystem. The median loss stands at $1 million, while the mean loss is calculated at $86 million, with a standard deviation of $1,598 million, all denominated in equivalent USD.

To further analyze fund loss, Figure 6 depicts the log-transformed severity (i.e., funds lost from an individual security incident) over the years. Remarkably, despite the rising frequency trend presented in Figure 1, there is a slight decrease in the severity of individual losses after 2020. A practical explanation of this observation may be the continuous maturation of existing developers and the increased robustness of training programs for developers who are entering the space.

Figure 6
THE LOG-TRANSFORMED SEVERITY EXPRESSED IN USD OVER THE YEARS

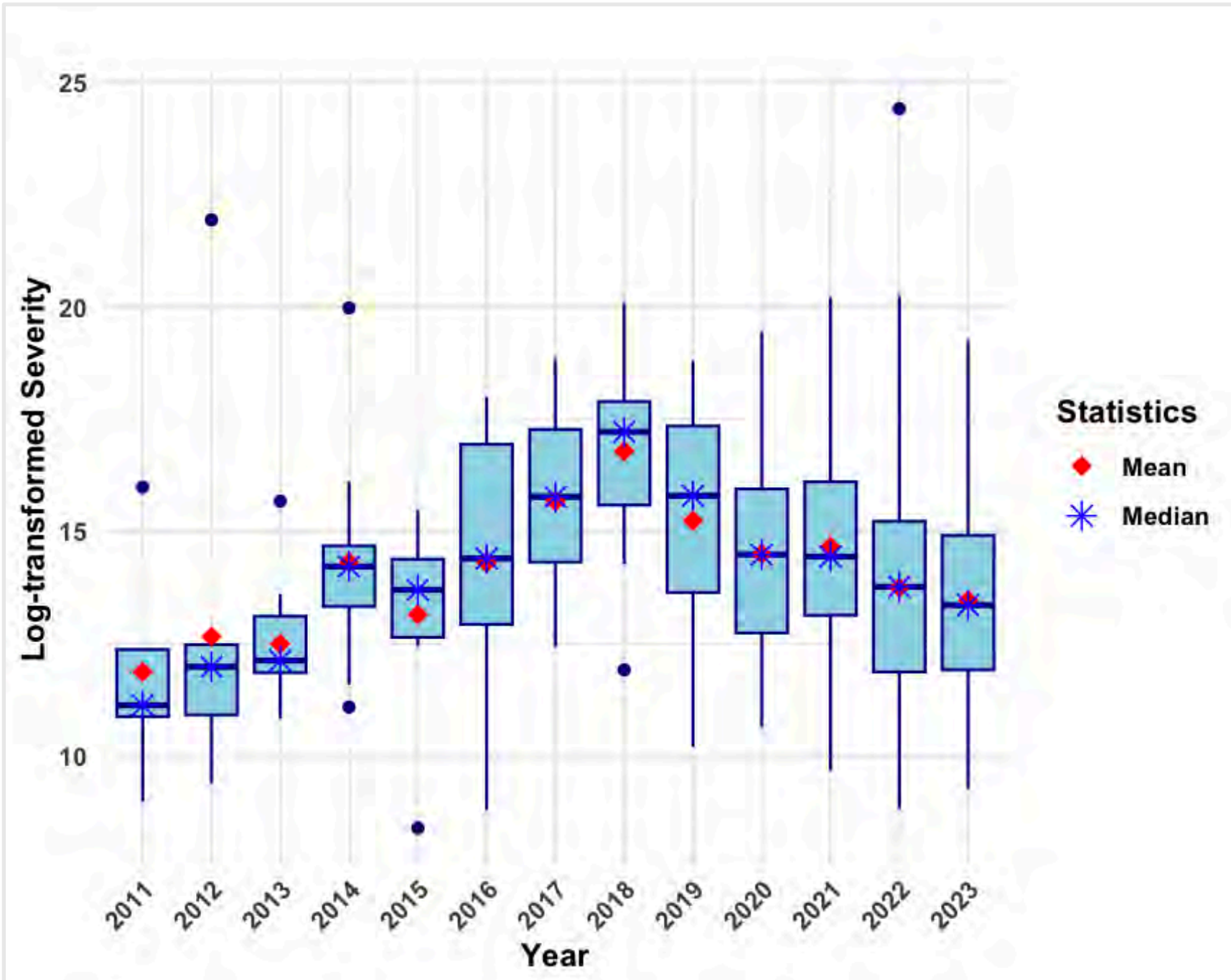

Figure 7 shows the total log-transformed severity of losses for each type of issue over the years. The data reveals a growing trend in losses due to access control issues, and despite only appearing after 2020, flash loan attacks have also led to significant losses. Excluding the “other” type of losses, all categories show a general increase. This trend suggests that hackers often target custom-built smart contracts rather than more established and thoroughly tested systems. In particular, direct exploits tend to occur more frequently as they target teams that venture beyond conventional approaches and push the boundaries of innovation. This represents a certain risk that comes with pioneering new technologies.

Figure 7

THE LOG-TRANSFORMED SEVERITY EXPRESSED IN USD ATTRIBUTED TO EACH TYPE OF ISSUE OVER THE YEARS

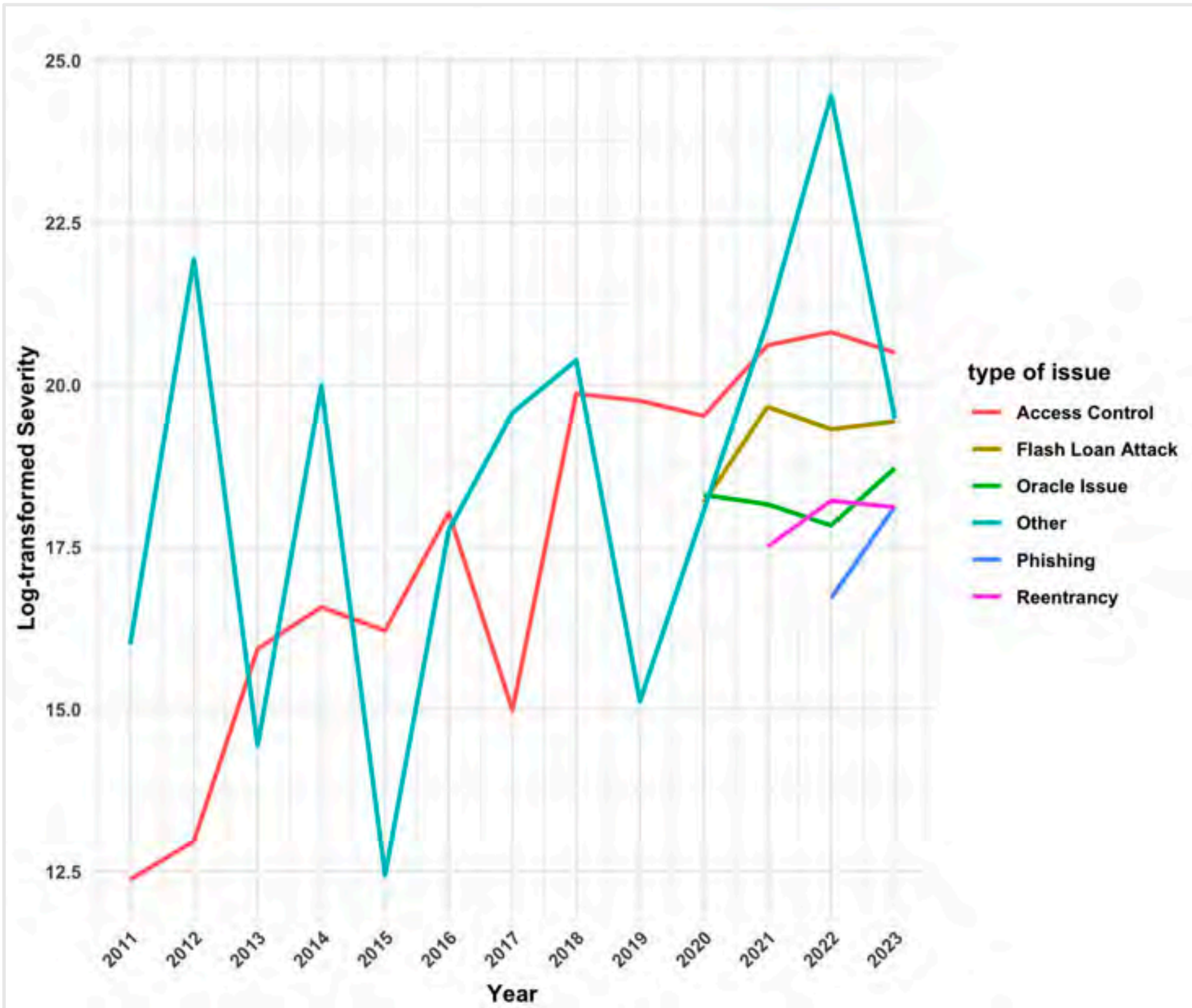

Figure 8 presents boxplots depicting the log-transformed severity for BSC, ETH, and other chains from 2020 to 2023, encompassing a total of 540 incidents. Interestingly, the other chains exhibit the highest mean and median values, whereas BSC displays the lowest mean and median.

Figure 8
BOXPLOTS ILLUSTRATING THE LOG-TRANSFORMED SEVERITY EXPRESSED IN USD WITHIN EACH CHAIN FROM 2020 TO 2023

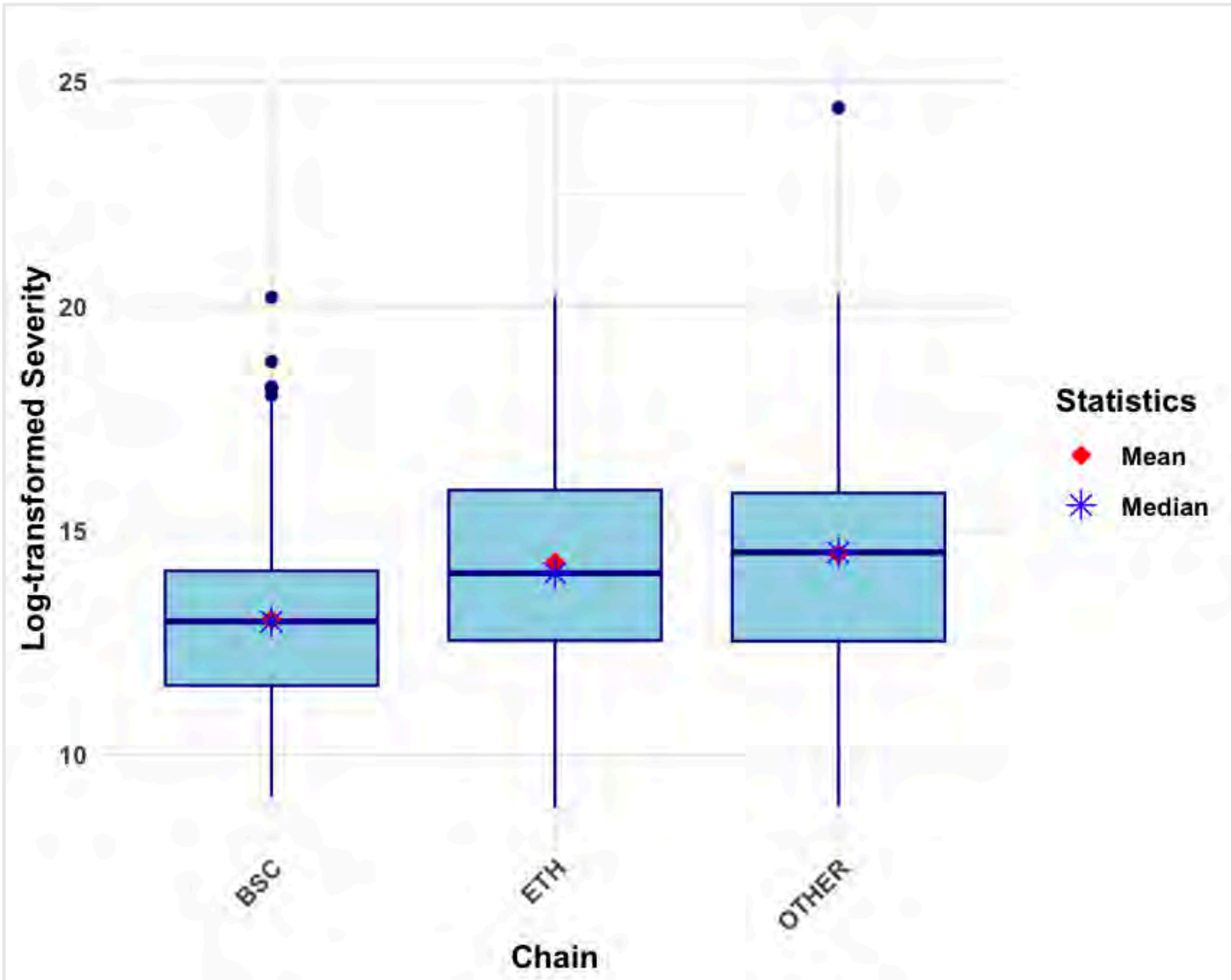

## Section 4: Loss Models for Pricing and Risk Management

Prospective digital asset insurers will rely on security risk models to support decision-making regarding pricing and risk management. In this section, our objective is to propose a class of practical models for quantifying the financial risk linked to security incidents.

Motivated by the observation that the majority of digital asset insurance buyers denominate coverage on a monthly basis, this paper centers on monthly security losses. We hypothesize that an insurance company offers cybersecurity insurance to a portfolio of $d$ protocols. At time $t$, let $L_{i,t}$ be the security loss random variable (RV) associated with the $i$-th protocol, $i = 1, \dots, d$, and $\mathbf{L}_t = (L_1, \dots, L_d)$ denotes the portfolio of losses which are dependent due to the prevalent practice of code sharing in the DeFi space. For illustrative purposes, this paper examines a portfolio comprising eight protocols, as outlined in Table 2. These protocols are selected due to their size and interconnectivity with the broader ecosystem.

Table 2
DESCRIPTION OF THE EIGHT PROTOCOLS IN THE ILLUSTRATIVE PORTFOLIO

| Protocol ID | Time of inception | Number of incidents | Chain | Protocol information |
|---|---|---|---|---|
| A | May, 2020 | 1 | ETH | An open source and non-custodial protocol to earn interest on deposits and borrow assets |
| B | Nov, 2018 | 2 | ETH | A decentralized exchange to trade all on-chain tokens |
| C | Feb, 2020 | 1 | ETH | A decentralized exchange focused on trading stable assets. |
| D | May, 2019 | 2 | ETH | An algorithmic interest rate protocol built for borrowing and lending |
| E | May, 2021 | 1 | ETH | A platform that optimizes yield strategies |
| F | Mar, 2020 | 3 | ETH | A decentralized automated market maker (AMM) protocol to create programmable liquidity. |
| G | Nov, 2020 | 3 | BSC | A decentralized marketplace for lenders and borrowers |
| H | Sep, 2021 | 1 | OTHER | A decentralized spot and perpetual exchange |

Further, let $N_{i,t}$ be the monthly security event frequency RV associated with the $i$-th protocol. Based on empirical evidence and typical policy designs, it's reasonable to assume that there's at most one security event occurring within a given month. Thereby, we set $N_{i,t} \sim Bernoulli(\pi^F_{i,t})$, where $\pi^F_{i,t} \in (0,1)$ is the incident probability of the $i$-th protocol at time $t$. Equivalently, we have

$$N_{i,t} = \begin{cases} 1 & with\ probability\ \pi^F_{i,t} \\ 0 & with\ probability\ 1 - \pi^F_{i,t} \end{cases}.$$

Thereby, the security loss RV can be written as

$$L_{i,t} = \mathbf{I}(N_{i,t} = 0) \times 0 + \mathbf{I}(N_{i,t} = 1) \times Y_{i,t},$$

where $Y_{i,t}$ denotes the loss severity assuming that there is a security event occurring at time $t$.

It's important to remind readers that due to common code-sharing practices in the digital asset ecosystem, potential security losses across different protocols may be interdependent. Specifically, if an attacker exploits a common vulnerability existing among multiple protocols, more than one protocol could be compromised in quick succession. This scenario poses a daunting challenge to insurers and threatens their financial stability. Therefore, in the context of risk management and capital calculation, it's crucial to account for the potential interdependence among individual protocol losses. Mainly, the dependence of $\mathbf{L}_t$ is induced by the dependence presenting among the frequency RV's, $\mathbf{N}_t = (N_{1,t}, \dots, N_{d,t})$. On the other hand, given the random nature of attacker behaviors across individual protocols, it is reasonable to assume that the loss severities among different protocols, $(Y_{1,t}, \dots, Y_{d,t})$, are independent.

In the following sections, we will outline the models we propose for studying the frequency and severity components. Given the evolving nature of the DeFi ecosystem, it's important to acknowledge the limitations stemming from the relatively limited availability of incident data. Therefore, we should explore creative solutions that strike a balance between statistical rigor and practical constraints.

### 4.1 MODELING THE FREQUENCY COMPONENT

Considering the discussions in Section 3, it's evident that the number of security incidents has significantly risen in recent years, corresponding to the continuous growth of digital assets. Therefore, we recommend modeling the attack frequency using the following logistic regression (McCullagh, 2019):

$$logit\left(\pi_{i,t}^{F}\right) = \alpha_{i,0} + \alpha_{i.1} \times log\left(TVL_{i,t}\right),$$

where $TVL_{i,t}$ denotes the asset value of the *i*-th protocol at time $t$. In our unpublished analysis, we explored alternative non-linear regression structures, and we concluded that the one mentioned above provides the best fit. We also explored the possibility of using the monthly growth rate of assets as a covariate. The underlying hypothesis is that protocols experiencing significant increases in assets may lack the necessary security infrastructure to accommodate such growth, thereby increasing their vulnerability to hacking incidents. However, our analysis suggests that, given the currently available data, utilizing the TVL directly as a covariate may be more effective than using its monthly growth rate.

To fit the logistic regression model mentioned above to the data, for a given protocol, we treat the historical monthly number of incident events as independent observations and then use the maximum likelihood method to estimate the regression coefficients. It is well-known that when the fitted probabilities of a logistic regression are very close to zero or one, convergence problems can occur for the gradient descent algorithm used to find the maximum likelihood estimate. This situation is often encountered in cyber security event modeling, where the frequency of security incidents is very low. When convergence problems arise, we can use penalized logistic regression techniques such as lasso or elastic-net to address the issue (James et al., 2013).

We utilize all available data until December 2023 as training data and aim to predict the attack probability for January 2024. Table 3 presents the coefficient estimates for the eight protocols included in the hypothetical portfolios. To aid interpretation, all covariate data were standardized to have a mean of zero and a standard deviation of one. The sign (positive or negative) of the coefficient indicates the direction of the relationship between the predictor variable and the log-odds of the response variable. The magnitude of the coefficient indicates the strength of the relationship between the predictor variable and the log-odds of the response variable, with larger magnitudes suggesting a stronger effect. Each coefficient represents the change in the log-odds of the response variable for a one-unit change in the predictor variable. As shown, most coefficient estimates are positive, indicating that higher TVL leads to a higher attack likelihood. However, a few protocols have slightly negative TVL coefficient estimates. A practical explanation is the simplicity and time-tested nature of these smart contracts prior to significant assets being accrued.

While the log TVL coefficients appear insignificant across all considered protocols, we believe that this is primarily due to the scarcity and sparsity of the data. Nonetheless, we maintain the inclusion of TVL as a predictor in the attack probability model, as it serves as a crucial vulnerability indicator. Protocols with higher TVL are likely to attract more attackers. Separately, another area of vulnerability pertains to the security measures implemented by individual protocols to mitigate exploit behaviors on an ongoing basis. A prospective insurer can combine industry standard monthly protocol with ongoing operational monitoring to develop risk scores and serve as useful predictors for loss analysis.

Table 3
SUMMARY OF THE REGRESSION COEFFICENT ESTIMATES AS WELL AS THE P-VALUES OF THE HL GOODNESS-OF-FIT TEST

| Protocol ID | $\hat{\alpha}_0$ | $\hat{\alpha}_1$ | P-value for the HL goodness-of-fit test |
|---|---|---|---|
| A | -3.7792***<br>(1.0318) | 0.2164<br>(1.2308) | 0.2653 |
| B | -3.4162***<br>(0.7287) | -0.1657<br>(0.6707) | 0.3702 |
| C | -4.0263**<br>(1.2740) | 0.7691<br>(1.7972) | 0.476 |
| D | -3.740**<br>(1.222) | 1.302<br>(1.877) | 0.686 |
| E | -3.4775**<br>(1.0583) | 0.2712<br>(1.0327) | 0.3655 |
| F | -2.7188 ***<br>(0.6113) | -0.2449<br>(0.4185) | 0.5611 |
| G | -3.1132**<br>(0.9518) | 1.5579<br>(1.0134) | 0.4418 |
| H | -3.831*<br>(1.891) | 2.149<br>(5.147) | 0.1710 |

To validate the adequacy of the proposed frequency model in capturing incident data trends, we employ the Hosmer-Lemeshow (HL) goodness-of-fit test (McCullagh, 2019). This test is designed to assess the following hypothesis:

$H_0$: The logistic regression model fits the observed data well;

v.s.

$H_1$: The logistic regression model does not fit the observed data well.

In simple terms, the HL test evaluates whether the observed and expected frequencies of the outcome variable are significantly different across different groups of the predictor variable, using the generalized Pearson chi-square statistic. The p-values of the HL goodness-of-fit test are displayed in the last column of Table 3. If P-value is larger than 0.05 or 0.1, depending on the model user's subjective preference, the logistic regression model could be considered good fit (i.e., failing to reject the null hypothesis $H_0$).

Finally, Table 4 displays the predicted attack probability based on the proposed frequency model.

Table 4
SUMMARY OF THE ATTACK PROBABILITY PREDICTIONS

| Protocol ID | Predicted attack probability |
|---|---|
| A | 2.4025% |
| B | 2.8789% |
| C | 1.6133% |
| D | 3.6861% |
| E | 2.1658% |
| F | 5.8898% |
| G | 4.6547% |
| H | 5.2114% |

It is important to note that some protocols in the market have never experienced attacks in their history. Therefore, the frequency model discussed here should not be directly applied to model these protocols. Instead, analysts can utilize protocols with similar business functions, contracts, and chain of deployment to construct interval approximations for those that have not yet experienced attacks.

## 4.2 FREQUENCY DEPENDENCE

As previously discussed, incident frequencies among different protocols within the portfolio are likely to be interdependent (Lee et al., 2023). However, rigorously, and statistically estimating these dependencies poses significant challenges due to data limitations. Specifically, the incident data are characterized by sparsity, with zero incident events occurring in most months. Among the incident data for the eight selected protocols, there is only one instance in which exploit events are observed among multiple protocols during the same month. Additionally, if we were to jointly estimate the occurrence of dependencies among all selected protocols, the training time window would need to encompass the existence of all selected protocols. However, some selected protocols have relatively short histories since their inception. Consequently, the available training data is limited, making it difficult to obtain reliable estimation results.

Acknowledging the criticality of capturing the interdependence of incident frequencies to assess the systemic risk within the digital asset portfolio, we advocate for the integration of a copula approach with engineering knowledge to tackle the aforementioned data limitation (Joe, 2014). Specifically, in order to obtain a reliable approximation of the dependence in incident frequencies, we resort to a similarity matrix, denoted by $\mathbf{\Psi}$ developed by the MetaRisk Labs team. This proprietary matrix incorporates factors including code similarity, commercial purpose, and operational overlap. The similarity matrix for the eight selected protocols is summarized in Table 5.

Table 5

SUMMARY OF THE SIMILARITY MATRIX FOR THE EIGHT SELECTED PROTOCOLS IN THE HYPOTHETICAL PORTFOLIO

| | A | B | C | D | E | F | G | H |
|---|---|---|---|---|---|---|---|---|
| A | 1.0 | 0.3 | 0.3 | 0.8 | 0.3 | 0.3 | 0.3 | 0.2 |
| B | 0.3 | 1.0 | 0.6 | 0.3 | 0.2 | 0.4 | 0.4 | 0.3 |
| C | 0.3 | 0.6 | 1.0 | 0.3 | 0.5 | 0.3 | 0.3 | 0.2 |
| D | 0.8 | 0.3 | 0.3 | 1.0 | 0.2 | 0.2 | 0.3 | 0.3 |
| E | 0.3 | 0.2 | 0.5 | 0.2 | 1.0 | 0.2 | 0.1 | 0.1 |
| F | 0.3 | 0.4 | 0.3 | 0.2 | 0.2 | 1.0 | 0.1 | 0.1 |
| G | 0.3 | 0.4 | 0.3 | 0.3 | 0.1 | 0.1 | 1.0 | 0.1 |
| H | 0.2 | 0.3 | 0.2 | 0.3 | 0.1 | 0.1 | 0.1 | 1.0 |

The aforementioned similarity matrix only provides information about pair-wise dependence among protocols. The elliptical copula is a useful mathematical framework for constructing high-dimensional dependence models, particularly when only pair-wise dependence is specified. Specifically, we use the similarity matrix as the correlation matrix in the Gaussian copula to inject dependencies among the incidence frequency RV's such that

$$P\left(N_{1,t} \le n_{1,t}, \dots, N_{d,t} \le n_{d,t}\right) = \Phi_d\left(\Phi^{-1}\left(P\left(N_{1,t} \le n_{1,t}\right)\right), \dots, \Phi^{-1}\left(P\left(N_{d,t} \le n_{d,t}\right)\right); \mathbf{\Psi}\right),$$

where $\Phi_d(\cdot,;\ \boldsymbol{\Sigma})$ denotes the joint cumulative distribution function (CDF) of the $d$-dimension standard normal distribution with correlation matrix $\boldsymbol{\Sigma}$, $\Phi^{-1}$ is the inverse of the standard normal CDF, and $n_{i,t} \in 0,1, i = 1, \dots, d$.

We acknowledge that the proposed copula approach may provide a (overly) conservative estimate for the dependencies in security incident occurrence, as the similarity matrix serves as a cautious proxy for the true, unknown correlation matrix underlying the dependent incident frequencies. However, given the lack of adequate data, we contend that the copula approach presented above remains a viable solution for addressing the dependencies among incident frequencies.

### 4.3 MODELING THE SEVERITY COMPONENT

Now, we turn to modeling the severity component $Y_{i,t}$. Due to the sparsity of the incidence loss data for a given protocol, our severity modeling approach acts as data borrowing from similar security events by implementing a generalized linear model (GLM) approach (McCullagh, 2019) over all the loss data observed in the entire DeFi ecosystem.

We focus our analysis on data spanning from 2020 to 2023 for two primary reasons: i) Recent incidents – given the dynamic nature of protocol markets, it is customary to concentrate on recent years to capture the most up-to-date trends. In fact, most of the incidents happened in this study period (i.e., 85.6%). ii) Emergence of new chains – certain blockchain networks, such as BSC, have only been established in recent years. In relation, the development of smart contracts and the creation of a DeFi ecosystem have only been recent phenomena, so a long historical study is inappropriate.

Motivated by the preliminary analysis discussed in Section 3.2, we incorporate chain, time, and TVL as covariates in the regression models. TVL data were sourced from DefiLlama one day before the attack. However, a notable portion of protocols lacks TVL data. Out of the 540 observations studied, 336 protocols entirely lack TVL data, and 23 protocols have a TVL of 0. For these cases, we make a simplifying assumption that lost funds are equivalent to TVL, hence assuming that all assets were lost after the attack.

In practical scenarios, the focus often lies on understanding the proportion of funds lost relative to the TVL within the protocol. As such, our modeling approach directly targets this specific quantity. Figure 9 shows the proportion of funds lost related to the TVL. It is clearly seen that a significant portion of protocols encountering incidents suffered a complete loss of funds. Based on this observation, we recommend employing a two-part model to accurately capture the proportion of funds lost (Frees et al., 2013).

Figure 9
HISTOGRAM OF PROPORTION OF FUNDS LOSS RELATED TO THE TVL WITHIN THE PROTOCOL

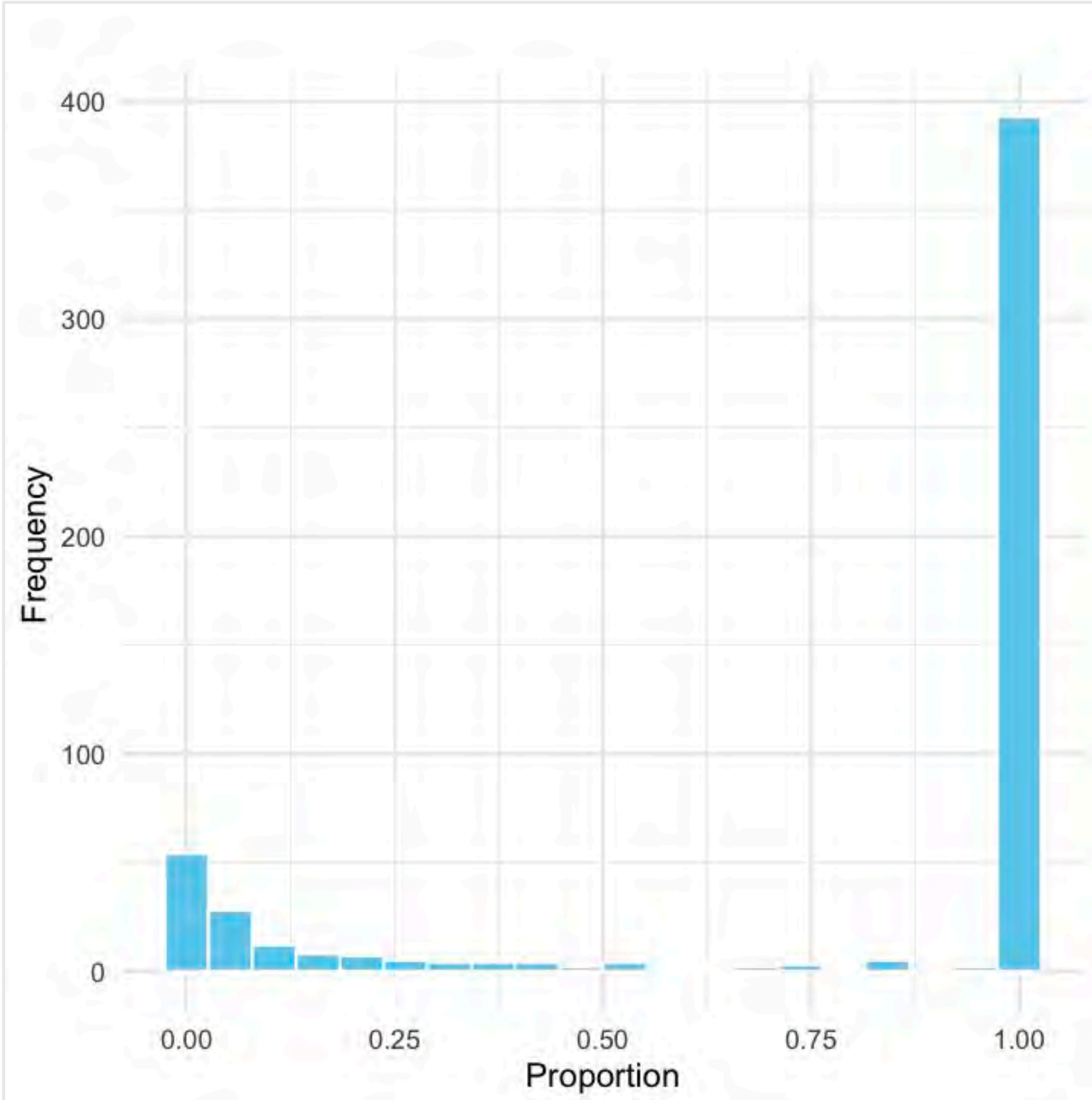

Specifically, the two-part model involves two steps: i) Determining whether a protocol incurs a total loss of funds. ii) Modeling the proportions of funds lost for protocols that do not experience a complete loss. Mathematically, let $R_{i,t} = Y_{i,t}/TVL_{i,t} \in (0,1]$, we model this loss ratio RV by

$$R_{i,t} = \left(1 - W_{i,t} \times R^*_{i,t} + W_{i,t}\right),$$

where $W_{i,t} \sim Bernoulli\left(\pi^S_{i,t}\right)$ indicates whether an attack causes a total fund loss, and $R^*_{i,t} \in (0,1)$ models the percentage of fund loss that is strictly smaller than one.

After the thorough model and variable selection processes, employing commonly used metrics such as Akaike Information Criterion (AIC) or Bayesian Information Criterion (BIC) (McCullagh, 2019), we recommend utilizing the following models:

$$logit\left(\pi^S_{i,t}\right) = \beta_0 + \beta_1 \times D_{ETH} + \beta_2 \times D_{OTHER} + \beta_3 \times log(TVL_{i,t}) + \beta_4 \times t + \beta_5 \times D_{ETH} \times t + \beta_6 \times D_{OTHER} \times t$$

and

$$logit(R_{i,t}^{*}) = \gamma_0 + \gamma_1 \times log(TVL_{I,T}) + \delta_{i,t}.$$

In the formulas above, $D_{ETH}$ and $D_{OTHER}$ are the dummy variables to indicate the chain type, and the residual $\delta_{i,t}$ is independently and identically distributed normal with mean 0 and variance $\sigma^2$. Table 6 summarizes the coefficient estimates of the above models.

Table 6

MODEL FITTINGS OF THE TWO-PART MODEL, WITH ** AND * REPRESENTING SIGNIFICANCE AT .05 AND .01 LEVELS, RESPECTIVELY

| | Model for total losses | Model for proportional losses |
|---|---|---|
| Intercept | 14.6015**<br>( 1.5747) | 10.0268**<br>(1.47145) |
| ETH | -2.9045*<br>( 1.1486) | -- |
| OTHER | -1.2515<br>( 1.3361) | -- |
| Log(TVL) | -0.6292**<br>(0.0599) | -0.7404**<br>(0.0824) |
| time | -1.3570**<br>( 0.3704) | -- |
| ETH:time | 1.1167**<br>(0.4103) | -- |
| OTHER:time | 0.2035<br>(.4716) | -- |

The following noteworthy observations have been made:

- Protocols with larger TVL are less likely to experience a complete loss of funds, suggesting a potential protective effect of higher TVL (e.g. audits, white hat hackers, investor due diligence, etc.).
- The passage of time demonstrates a negative correlation with the likelihood of total fund loss, possibly indicative of ongoing advancements in protocol security measures.
- Protocols associated with the ETH chain exhibit an increased probability of total fund loss over time. However, the primary effect of the ETH chain is negative, which could be attributed to the predominance of protocols within this chain category.

To validate the adequacy of the proposed model, we perform the HL goodness-of-fit test for the total loss model. A p-value of 0.4475 is obtained, indicating a good fit of the logistic regression to the observed total loss data. Regarding the proportional loss model, we examine the QQ-plot of the quantile residuals, as summarized in Figure 10. The QQ-plot demonstrates satisfactory fitting performance. The estimation of fund loss can be calculated by multiplying this proportion by the current TVL.

Figure 10

QQ-PLOT OF QUANTILE RESIDUALS OF THE FITTING OF THE PROPORTIONAL LOSS MODEL

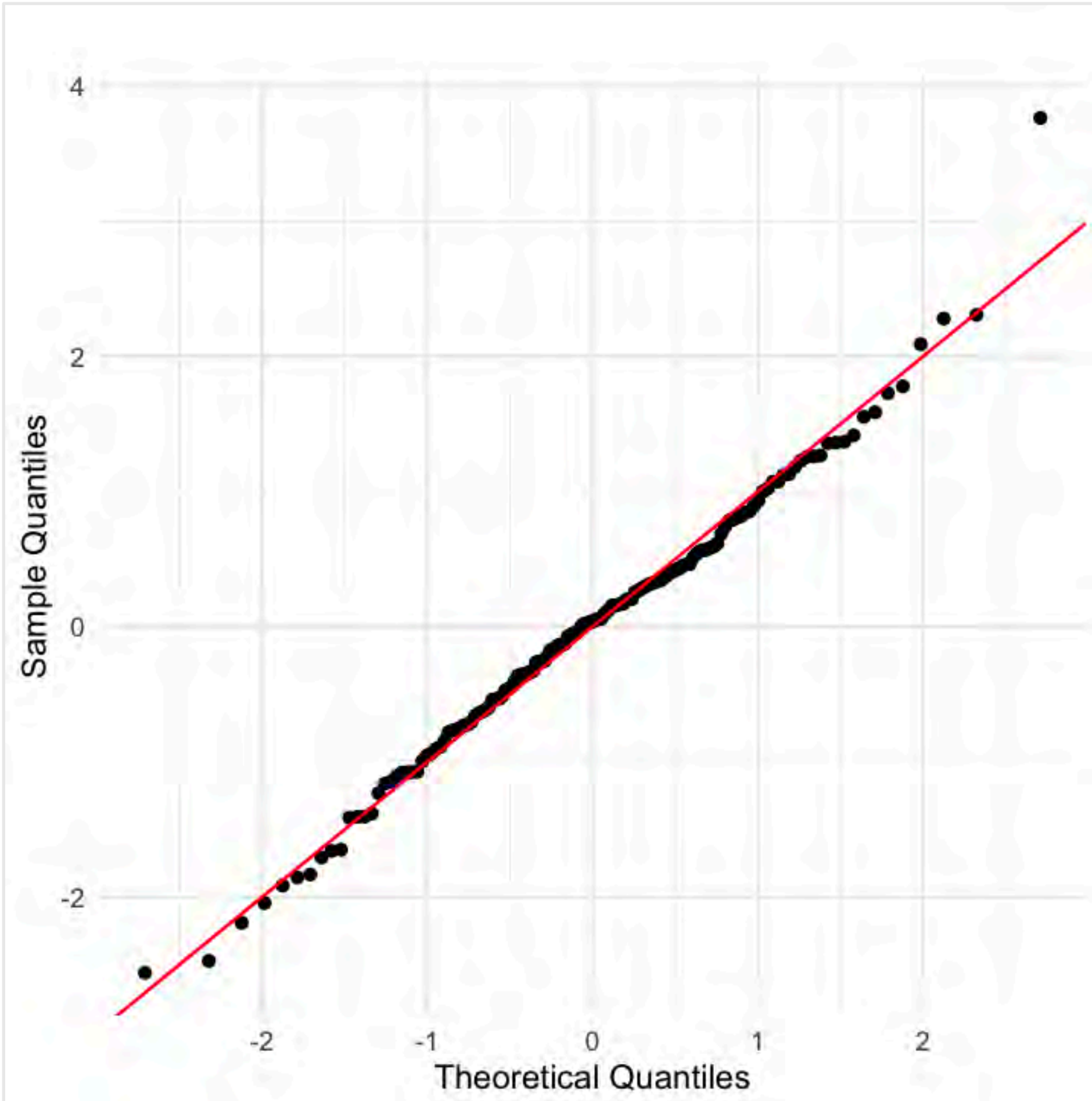

## Section 5: Insurance Applications

This section offers a simplified demonstration of the cyber security loss models outlined in Section 4. We explore two applications: pricing and tail risk management. We continue to examine the hypothetical portfolio detailed in Table 2. Additionally, we assume the existence of a sufficiently large insurer in the market capable of providing coverage for all TVL within the considered protocols. In practice, digital asset insurers may only cover a portion of the TVL, in which case premium calculations can be adjusted proportionally. For simplicity, we do not account for deductibles, co-insurance, and limits, although these can be easily incorporated into the proposed framework. To price cyber security for a given protocol, we consider two commonly used pricing principles:

- Expectation principle: $\rho_1(L_{i,t+1}) = (1+\theta) \times E(L_{i,t+1})$;
- Standard deviation principle: $\rho_2(L_{i,t+1}) = E(L_{i,t+1}) + \theta \times SD(L_{i,t+1})$.

In the formula above, $\theta > 0$ is the loading parameter, which reflects the risk preference of the insurer[7].

Capitalizing on the frequency and severity models laid out in Section 4, we can compute the quantities in the two pricing principles in the following manner:

$$E(L_{i,t+1}) = E(N_{i,t+1}) \times E(Y_{i,t+1}),$$

where

$$E(N_{i,t+1}) = \hat{\pi}^F_{i,t+1}$$

and

$$E(Y_{i,t+1}) = TVL_{i,t+1} \times \left[(1-\pi^S_{i,t+1}) \times E(R^*_{i,t+1}) + \pi^S_{i,t+1}\right].$$

Here, $\pi^F_{i,t+1}$ and $\pi^S_{i,t+1}$ are the predicted probability parameters computed based on the regression coefficient estimates outlined in Tables 3 and 6. To compute $E(R^*_{i,t+1})$, let us write

$$E(R^*_{i,t+1}) = E\left(logit^{-1}(\gamma_0 + \gamma_1\, log(TVL_{i,t+1}) + \delta_{i,t+1})\right),$$

where $\delta_{i,t+1}$ is independently distributed normal with mean 0 and variance $\hat{\sigma}^2$, the dispersion parameter estimate obtained in the estimation of the proportional loss model. To the best of our knowledge, there is no closed formula available to compute the expectation above. Therefore, we resort to Monte Carlo simulation to approximate the expectation $E(R^*_{i,t+1})$.

To calculate the standard deviation principle, we further need

$$SD(L_{i,t+1}) = \sqrt{Var(L_{i,t+1})} = \sqrt{E(N_{i,t+1}) \times E(Y^2_{i,t+1}) - E(N_{i,t+1})^2 \times E(Y_{i,t+1})^2}.$$

In the calculation of $E(L_{i,t})$, we have already computed $E(N_{i,t+1})$ and $E(Y_{i,t+1})$. The remaining term $E(Y^2_{i,t+1})$ is computed via Monte Carlo simulation in a similar manner to the calculation of $E(Y_{i,t+1})$.

Table 7 displays the premium calculation for the eight protocols involved in the hypothetical portfolio. We observe that there is a notable discrepancy in premium amounts between the expectation principle and the standard deviation principle for each protocol. Premiums calculated under the standard deviation principle tend to be substantially higher compared to those under the expectation principle, highlighting the large deviation and heavy tail natures associated with cyber security losses. Different protocols exhibit varying levels of perceived risk, as evidenced by the differences in premium amounts. Protocols such as F, H, and D have relatively higher premium rates compared to others, suggesting they may face greater cyber security risks or have characteristics that contribute to higher potential losses. Protocols with higher premiums may attract attention from developers and stakeholders seeking to enhance cyber security measures and mitigate risks. Lowering the perceived risk profile of these protocols could lead to reductions in insurance premiums over time.

---

[7] Throughout the numerical analysis, we set $\theta = 0.5$ for illustration.

Table 7
SUMMARY OF THE PREMIUM CALCULATIONS

| Protocol ID | Predicted attack probability | Predicted loss percentage | Expectation principle (% protocol TVL) | SD principle (% protocol TVL) |
|---|---|---|---|---|
| A | 2.4025% | 4.3901% | 10,800,265<br>0.1585% | 101,496,825<br>1.4898% |
| B | 2.8789% | 5.9181% | 10,372,064<br>0.2552% | 78,987,279<br>1.9434% |
| C | 1.6133% | 8.8469% | 4,255,216<br>0.2136% | 35,684,547<br>1.7911% |
| D | 3.6861% | 8.5272% | 10,059,195<br>0.4721% | 58,818,128<br>2.7603% |
| E | 2.1658% | 9.9399% | 5,222,052<br>0.3237% | 36,398,644<br>2.2562% |
| F | 5.8898% | 13.4714% | 10,901,942<br>1.1920% | 43,425,138<br>4.7481% |
| G | 4.6547% | 6.0779% | 4,205,261<br>0.4240% | 21,826,721<br>2.2008% |
| H | 5.2114% | 7.1940% | 3,326,761<br>0.5626% | 15,065,769<br>2.5480% |

To demonstrate the risk management application of the proposed model, we focus on tail risk assessment for the portfolio. To this end, let $S = L_{1,t+1} + \cdots + L_{d,t+1}$ be the aggregate risk across the portfolio. We consider two state-of-the-art risk measures, namely the Value-at-Risk (VaR):

$$VaR_q(S) = inf\{s \in R: F_S(s) \geq q\}, \quad q \in (0,1)$$

and the conditional tail expectation (CTE):

$$CTE_q(S) = E(S|S > VaR_q(S)).$$

There are no simple closed-form formulas for computing the two aforementioned risk measures. We utilize Monte Carlo simulation to calculate these risk measures, which are summarized in Table 8. When the confidence level is close to one, we observe that both risk measures are higher under the dependence scenario, emphasizing the significance of considering the simultaneous occurrence of security events in managing a digital asset insurance portfolio.

Table 8
SUMMARY OF RISK MEASURE CALCULATIONS WITH CORRESPONDING PERCENTAGES OUT OF TOTAL ASSETS AMONG EIGHT SELECTED PROTOCOLS

| Confidence level | VaR computed with dependence | VaR computed without dependence | CTE computed with dependence | CTE computed without dependence |
|---|---|---|---|---|
| 90% | 12,701,792<br>0.0665% | 16,236,219<br>0.0850% | 390,787,994<br>2.0449% | 388,393,907<br>2.0324% |
| 95% | 78,548,729<br>0.4110% | 83,170,666<br>0.4352% | 746,914,073<br>3.9084% | 737,902,839<br>3.8612% |
| 99% | 930,857,773<br>4.8709% | 916,596,940<br>4.7963% | 2,366,912,586<br>12.3854% | 2,307,669,732<br>12.0754% |

## Section 6: Conclusion and Discussion

Our study has uncovered several pivotal insights. Firstly, we've noted that protocols with larger asset bases have a lower likelihood of experiencing complete fund loss, suggesting a protective effect potentially caused by more rigorous audits, security community engagement, and enhanced investor due diligence. Additionally, our findings indicate a negative correlation between the passage of time and the probability of total fund loss, hinting at continuous improvements in protocol security. However, protocols have shown an increased probability of loss as time progresses, highlighting complex, evolving risk dynamics within the digital asset ecosystem.

This report scratches the surface of the intricate interplay between innovation and risk in digital finance. While we provide initial tools and insights, there is a vast, uncharted territory that beckons for deeper exploration. For forward-looking insurers, the potential for generating outsized, uncorrelated returns compared to traditional premium sources remains substantial as new ventures develop and innovations such as built-in contract protection emerge. Thriving in this burgeoning environment requires a greater understanding of blockchain risks and the development of sophisticated strategies to evaluate, monitor, and mitigate potential losses.

We stand at the brink of an evolving era in digital asset risk management. Our initial findings have carved out a basic framework for insurers, offering a glimpse into the transformative potential of blockchain technology in reshaping risk assessment and security. However, as comprehensive as our explorations have been, they merely scratch the surface of what is possible. Looking ahead, the challenge for insurers is not only to adopt this framework but also to innovate upon it, continuously adapting to the ever-changing digital landscape. What new strategies will emerge as blockchain becomes an integral part of financial operations worldwide? This question leaves us poised for a future rich with unexplored opportunities.

## Section 7: Acknowledgments

The researchers' deepest gratitude goes to those without whose efforts this project could not have come to fruition: the Project Oversight Group and others for their diligent work overseeing questionnaire development, analyzing and discussing respondent answers, and reviewing and editing this report for accuracy and relevance.

Project Oversight Group members:

Min Ji, FSA, FIA, MAAA

Bernice Lim, FSA, CERA, FCIA, CFA

Monojit Samanta, FSA, FCIA

Feng Sun, FSA, CERA

Maggie Sun, ASA, MAAA

William Wilkins, ASA, CERA, FCAS, MAAA

At the Society of Actuaries Research Institute:

Rob Montgomery ASA, MAAA, FLMI, Consultant - Research Project Manager

The Society of Actuaries Research Institute would like to acknowledge the generous contribution of the Casualty Actuarial Society to the funding of this research.

## References

Caporale, G. M., Kang, W.-Y., Spagnolo, F., and Spagnolo, N. (2021). Cyber-attacks, spillovers and contagion in the cryptocurrency markets. *Journal of International Financial Markets, Institutions and Money*, 74:101298.

Chen, X., Liao, P., Zhang, Y., Huang, Y., and Zheng, Z. (2021). Understanding code reuse in smart contracts. In *2021 IEEE International Conference on Software Analysis, Evolution and Reengineering (SANER)*, pages 470–479. IEEE.

Danielius, P., Stolarski, P., & Masteika, S. (2020). Vulnerabilities and excess gas consumption analysis within ethereum-based smart contracts for electricity market. *In Business Information Systems Workshops*: BIS 2020 International Workshops, Colorado Springs, CO, USA, June 8–10, 2020, Revised Selected Papers 23, pages 99-110. Springer International Publishing.

Frees, E. W., Jin, X., & Lin, X. (2013). Actuarial applications of multivariate two-part regression models. *Annals of Actuarial Science*, 7(2), 258-287.

Gudgeon, L., Perez, D., Harz, D., Livshits, B., and Gervais, A. (2020). The decentralized financial crisis. In *2020 crypto valley conference on blockchain technology* (CVCBT), pages 1–15. IEEE.

He, R., Jin, Z., & Li, J. S. H. (2024). Modeling and management of cyber risk: a cross-disciplinary review. *Annals of Actuarial Science*, 1-40.

James, G., Witten, D., Hastie, T., & Tibshirani, R. (2013). *An Introduction to Statistical Learning*. New York: Springer.

Joe, H. (2014). *Dependence Modeling with Copulas*. CRC press.

Kwock, A., Lie, E., Weng, G., and Zhang, R. (2022). Decentralized insurance alternatives: Market landscape, opportunities, and challenges. *The SOA Research Institute*, pages 1–41.

Lee, S. S., Murashkin, A., Derka, M., & Gorzny, J. (2023). Sok: Not quite water under the bridge: Review of cross-chain bridge hacks. In 2023 *IEEE International Conference on Blockchain and Cryptocurrency* (ICBC), pages 1-14.

McCullagh, P. (2019). *Generalized Linear Models*. Routledge.

Qin, K., Zhou, L., Livshits, B., and Gervais, A. (2021). Attacking the defi ecosystem with flash loans for fun and profit. In *Financial Cryptography and Data Security*: 25th International Conference, FC 2021, Virtual Event, March 1–5, 2021, Revised Selected Papers, Part I, pages 3–32. Springer.

Sun, H., Xu, M., & Zhao, P. (2021). Modeling malicious hacking data breach risks. *North American Actuarial Journal*, 25(4), 484-502.

Sun, H., Xu, M., & Zhao, P. (2023). A multivariate frequency-severity framework for healthcare data breaches. *The Annals of Applied Statistics*, 17(1), 240-268.

Wan, Z., Xia, X., Lo, D., Chen, J., Luo, X., and Yang, X. (2021). Smart contract security: A practitioners' perspective. 2021 IEEE/ACM 43rd *International Conference on Software Engineering* (ICSE), pages 1410–1422.

Zhou, L., Xiong, X., Ernstberger, J., Chaliasos, S., Wang, Z., Wang, Y., Qin, K., Wattenhofer, R., Song, D., and Gervais, A. (2023). Sok: Decentralized finance (defi) attacks. In *2023 IEEE Symposium on Security and Privacy (SP),* pages 2444-2461. IEEE.

## About The Society of Actuaries Research Institute

Serving as the research arm of the Society of Actuaries (SOA), the SOA Research Institute provides objective, data-driven research bringing together tried and true practices and future-focused approaches to address societal challenges and your business needs. The Institute provides trusted knowledge, extensive experience and new technologies to help effectively identify, predict and manage risks.

Representing the thousands of actuaries who help conduct critical research, the SOA Research Institute provides clarity and solutions on risks and societal challenges. The Institute connects actuaries, academics, employers, the insurance industry, regulators, research partners, foundations and research institutions, sponsors and non-governmental organizations, building an effective network which provides support, knowledge and expertise regarding the management of risk to benefit the industry and the public.

Managed by experienced actuaries and research experts from a broad range of industries, the SOA Research Institute creates, funds, develops and distributes research to elevate actuaries as leaders in measuring and managing risk. These efforts include studies, essay collections, webcasts, research papers, survey reports, and original research on topics impacting society.

Harnessing its peer-reviewed research, leading-edge technologies, new data tools and innovative practices, the Institute seeks to understand the underlying causes of risk and the possible outcomes. The Institute develops objective research spanning a variety of topics with its strategic research programs: aging and retirement; actuarial innovation and technology; mortality and longevity; diversity, equity and inclusion; health care cost trends; and catastrophe and climate risk. The Institute has a large volume of topical research available, including an expanding collection of international and market-specific research, experience studies, models and timely research.

Society of Actuaries Research Institute
8770 W Bryn Mawr Ave, Suite 1000
Chicago, IL 60631
www.SOA.org

## About The Casualty Actuarial Society

The Casualty Actuarial Society (CAS) is a leading international organization for credentialing, professional education and research. Founded in 1914, the CAS is the world's only actuarial organization focused exclusively on property-casualty risks and serves over 10,000 members worldwide. CAS members are sought after globally for their insights and ability to apply analytics to solve insurance and risk management problems.

As the world's premiere P&C actuarial research organization, the CAS reaches practicing actuaries across the globe with thought-leading concepts and solutions. The CAS has been conducting research since its inception. Today, the CAS provides thousands of open-source research papers, including its prestigious publication, *Variance* — all of which advance actuarial science and enhance the P&C insurance industry. Learn more at casact.org.

The Casualty Actuarial Society
4350 N. Fairfax Drive, Suite 250
Arlington, VA  22203
www.casact.org